# 1 Overview of Web Content Mining Tools

## Abdelhakim Herrouz
*Department of Computer Science, University Kasdi Merbah of Ouargla, Algeria*
Abdelhakim.herrouz@gmail.com

## Chabane Khentout
*Laboratoire des Réseaux et des Systèmes Distribués,*
*University Ferhat Abbas of Sétif, Algeria*
khentout@yahoo.fr

## Mahieddine Djoudi
*Department XLIM-SIC UMR CNRS 7252 & TechNE Research Group, University of Poitiers, Teleport 2,*
*Boulevard Marie et Pierre Curie, B.P 30179, 86960 Futuroscope Cedex, France*
mahieddine.djoudi@univ-poitiers.fr

-------------------------------------------------------ABSTRACT--------------------------------------------------------------
*Nowadays, the Web has become one of the most widespread platforms for information change and retrieval. As it becomes easier to publish documents, as the number of users, and thus publishers, increases and as the number of documents grows, searching for information is turning into a cumbersome and time-consuming operation. Due to heterogeneity and unstructured nature of the data available on the WWW, Web mining uses various data mining techniques to discover useful knowledge from Web hyperlinks, page content and usage log. The main uses of web content mining are to gather, categorize, organize and provide the best possible information available on the Web to the user requesting the information. The mining tools are imperative to scanning the many HTML documents, images, and text. Then, the result is used by the search engines. In this paper, we first introduce the concepts related to web mining; we then present an overview of different Web Content Mining tools. We conclude by presenting a comparative table of these tools based on some pertinent criteria.*

*Keywords - Structured Data Tools, Web, Web Content Mining, Web Mining.*

--------------------------------------------------------------------------------------------------------------------------------
Date of Submission: 05 June 2013, 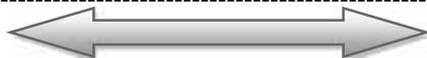 Date of Publication: 15 June 2013
--------------------------------------------------------------------------------------------------------------------------------

## I. INTRODUCTION

Due to the rapid growth of the Web, sites appear and disappear, content is modified and it becomes impossible to master their organization. The nature of the environment itself imposes some disadvantages: Internet is a network of worldwide level, constantly changing and non-structured [1]. The Web is the largest data source in the world. Web mining aims to extract and mine useful knowledge from the Web. It is a multi-disciplinary field involving data mining, machine learning, natural language processing, statistics, databases, information retrieval, multimedia, etc.

The Web offers an unprecedented opportunity and challenge to data mining [2]:

- The amount of information on the Web is huge, and easily accessible.
- The coverage of Web information is very wide and diverse. One can find information about almost anything.
- Information/data of almost all types exist on the Web, e.g., structured tables, texts, multimedia data, etc.
- Much of the Web information is semi-structured due to the nested structure of HTML code.
- Much of the Web information is linked. There are hyperlinks among pages within a site, and across different sites.
- Much of the Web information is redundant. The same piece of information or its variants may appear in many pages.

- The Web is noisy. A Web page typically contains a mixture of many kinds of information, e.g., main contents, advertisements, navigation panels, copyright notices, etc.
- The Web is also about services. Many Web sites and pages enable people to perform operations with input parameters, i.e., they provide services.
- The Web is dynamic. Information on the Web changes constantly. Keeping up with the changes and monitoring the changes are important issues.
- Above all, the Web is truly a virtual society. It is not only about data, information and services, but also about interactions among people, organizations and automatic systems.

These characteristics present both challenges and opportunities for mining and discovery of information and knowledge from the Web [3]. The conclusion from the analysis of these features and problems is certainly to develop tools for the Web Mining field.

The remaining sections of the paper are organized as follows. First we introduce the general concepts related to Web mining; we then present an overview of different Web Content Mining tools available in the literature. After that, we conclude by presenting a comparative table of these tools based on some pertinent criteria.

## II.  WEB MINING

Data mining is commonly defined as the process of discovering useful patterns or knowledge from data sources (e.g., databases, texts, images, the Web, etc.). The patterns must be valid, potentially useful, and understandable [3].

Generally, data mining uses structured data stored in relational tables, spread sheets, or flat files in the tabular form. With the growth of the Web and text documents, Web mining and text mining are becoming Increasingly important and popular.

Web mining aims to discover useful information or knowledge from the Web hyperlink structure, page content, and usage data. Zdravko and Daniel define web mining as follows: "For web mining, we refer to the application of data mining methodologies, techniques, and models to the variety of data forms, structures, and usage patterns that comprise the World Wide Web "[4]. In general, Web mining can be divided into three separate categories depending on the type of data to explore: Web Structure Mining,Web Content Mining and Web Usage Mining. It can be categorized in the fig.1 [5].

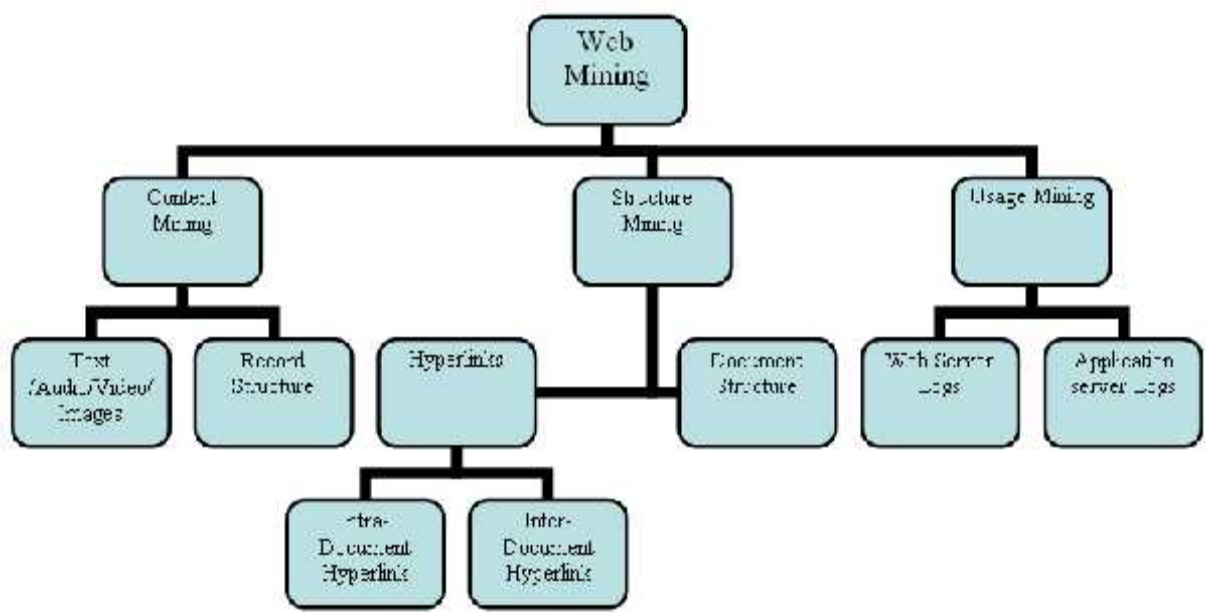

**Fig. 1:** Web Mining Taxonomy

**Web structure mining:** It consists to analyze how the pages are written. It discovers useful knowledge from hyperlinks, which represent the structure of the Web. For example, from the links, we can discover important

Web pages, which, incidentally, is a key technology used in search engines. We can also discover communities of users who share common interests. Traditional data mining does not perform such tasks because there is usually no link structure in a relational table.

**Web content mining:** It aims to extract information relating to the website page contents. It extracts or mines useful information or knowledge from Web page contents. For example, we can automatically classify and cluster Web pages according to their topics. These tasks are similar to those in traditional data mining. However, we can also discover patterns in Web pages to extract useful data such as descriptions of products, postings of forums, etc, for many purposes. Furthermore, we can mine customer reviews and forum postings to discover consumer sentiments. These are not traditional data mining tasks.

**Web usage mining:** It is also known as Web log mining, is used to analyze the behavior of website users. It refers to the discovery of user access patterns from Web usage logs, which record every click made by each user. Web usage mining applies many data mining algorithms. One of the key issues in Web usage mining is the pre-processing of clickstream data in usage logs in order to produce the right data for mining [3].

## III. WEB CONTENT MINING

This category is performed by extracting useful information from the content of a web page/site. It includes extraction of structured data/information from web pages, identification, match, and integration of semantically similar data, opinion extraction from online sources, and concept hierarchy, ontology, or knowledge integration [6].

Web content mining identifies the useful information from the Web Contents/data/documents. However, such a data in its broader form has to be further narrowed down to useful information. The web content data consist of structured data such as data in the tables, unstructured data such as free texts, and semi-structured data such as HTML documents. Two main approaches are used in Web Content Mining: (1) Unstructured text mining approach and (2) Semi-Structured and Structured mining approach [7].

**Unstructured Text Data Mining:** Web content data is much of unstructured text data. The research around applying data mining techniques to unstructured text is termed Knowledge Discovery in Texts (KDT), or text data mining, or text mining. Hence, one could consider text mining as an instance of Web content mining. To provide effectively exploitable results, preprocessing steps for any structured data is done by means of information extraction, text categorization, or applying NLP techniques.

**Semi-Structured and Structured Data Mining:** Structured data on the Web are often very important as they represent their host pages, due to this reason it is important and popular. Structured data is also easier to extract compared to unstructured texts. Semi-structured data is a point of convergence for the Web and database communities: the former deals with documents, the latter with data. The form of that data is evolving from rigidly structured relational tables with numbers and strings to enable the natural representation of complex real-world objects like books, papers, movies, etc., without sending the application writer into contortions. Emergent representations for semi-structured data (such as XML) are variations on the Object Exchange Model (OEM). In OEM, data is in the form of atomic or compound objects: atomic objects may be integers or strings; compound objects refer to other objects through labeled edges. HTML is a special case of such intra-document structure.

## IV. WEB CONTENT MINING TOOLS OVERVIEW

With the flood of information and data on the Web, the content mining tools helps to download the essential information that one would require. They collect appropriate and perfectly fitting information [7], [8], [9]. Some of them are Screen-scraper, Automation Anywhere 6.1, Web Info Extractor, Mozenda, and Web Content Extractor.

**Screen-scaper** [10]: Screen-scraping is a tool for extracting/mining information from web sites. It can be used for searching a database, SQL server or SQL database, which interfaces with the software, to achieve the content mining requirements. The programming languages like Java, .NET, PHP, Visual Basic and Active Server Pages (ASP) can also be used to access screen scraper.

Features: Screen-scraper present a graphical interface allowing the user to designate URL's, data elements to be extracted and scripting logic to traverse pages and work with mined data. Once these items have been created, from external languages such as .NET, Java, PHP, and Active Server Pages, Screen-scraper can be invoked. This also facilitates scraping of information at periodic intervals. One of the most regular usages of this software and services is to mine data on products and download them to a spreadsheet. A classier example would be a meta-search engine where in a search query entered by a user is concurrently run on multiple web sites in real-time, after which the results are displayed in a single interface.

**Automation Anywhere 6.1 (AA)** [11]: AA is a Web data extraction tool used for retrieving web data, screen scrape from Web pages or use it for Web mining.
Features:

- Unique SMART Automation Technology for fast automation of complex tasks.
- Record keyboard and mouse or use point and click wizards to create automated tasks quickly.
- Web record and Web data extraction.
- 305 plus actions are included: Internet, conditional, loop, prompt, file management, database and system, automatic email notifications, task chaining, etc.

**Web Info Extractor (WIE)** [12]: This is a tool for data mining, extracting Web content, and Web content analysis. WIE can extract structured or unstructured data from Web page, reform into local file or save to database, place into Web server.
Features:

- No need to learn boring and complex template rules, and it is easy to define extract tool.
- Extract tabular as well as unstructured data to file or database.
- Monitor Web pages and extract new content when update.
- Can deal with text, image and other link file.
- Can deal with Web page in all language.
- Running multi-task at the same time.
- Support recursive task definition.

**Mozenda** [13]: This tool enables users to extract and manage Web data. Users can setup agents that routinely extract, store, and publish data to multiple destinations. Once information is in Mozenda systems, users can format, repurpose, and mashup the data to be used in other applications or as intelligence. There are two parts of Mozenda's scaper tool:

- Mozenda Web Console: It is a Web application that allows user to run agents, view & organize results, and export publish data extracted.
- Agent Builder: It is a Windows application used to build data extraction project.
  Features:
- Easy to use.
- Plateform independency. However, Mozenda Agent Builder only runs on Windows.
- Working place independence: Tune the scraper, manage the scraping process and get scraped data from any computer connected to the Web.

**Web Content Extractor (WCE)** [14]: WCE is a powerful and easy to use data extraction tool for Web scraping, data mining or data extraction from the Internet. It offers a friendly, wizard-driven interface that will help through the process of building a data extraction pattern and creating crawling rules in a simple point-and-click manner. This tool allows users to extract data from various websites such as online stores, online auctions, shopping sites, real estate sites, financial sites, business directories, etc. The extracted data can be exported to a variety of formats, including Microsoft Excel (CSV), Access, TXT, HTML, XML, SQL script, MySQL script and to any ODBC data source.
Features:

- Helps to extract/collect the market figures, product pricing data, or real estate data.
- Helps users to extract the information about books, including their titles, authors, descriptions, ISBNs, images, and prices, from online book sellers.
- Assists users in automate extraction of auction information from auction sites.
- Assists to Journalists extract news and articles from news sites.
- Helps people seeking a job extract job postings from online job websites. Find a new job faster and with minimum inconveniences
- Extract the online information about vacation and holiday places, including their names, addresses, descriptions, images, and prices, from web sites.

## V. COMPARATIVE STUDY

**Comparison Criteria:** The different Web Content Mining tools are difficult to compare because of the variety of goals and contexts. We compare the existing tools based on the following four points:

- Usability (User friendly).
- Possibility to Record the Data.
- Perform on Structured web Data.
- Perform on Unstructured web Data.

**Comparative Table:** The following table summarizes the characteristics of these Content Mining Tools. In the columns, we use the following symbols:

- - : for No
- √: for Yes
- U: Usability
- R.D: Record the Data
- E.S.D: Extract Structured web Data
- E.U.D: Extract Unstructured web Data

Table 1: Comparative Table of Web Content Tools

| Tool | U | R.D | E.S.D | E.U.D |
|------|---|-----|-------|-------|
| Screen-Scraper | - | - | | |
| AA 6.1 | | | | |
| WIE | | - | | |
| Mazenda | | - | | |
| WCE | - | - | | |

As can be seen from the table above, we can notice the following:

- All the tools automate the business task and retrieve the web data in an efficient way.
- All the tools are performed on structured and unstructured web data.
- AA 6.1 allows recording of actions. This facility unique and it is not provided in the other tools.

Screen-scrapper needs prior knowledge of proxy server and some knowledge of HTML and HTTP where as other tools do not require any such knowledge and it need Internet connection to run.

## VI. CONCLUSION

The Web Data Mining tools are primordial to scanning the many HTML documents, images, and text provided on Web pages. The result is provided to the search engines, in order of relevance giving more productive results of each search. In this paper, we presented a non-exhaustive list of the available Web Content Mining Tools. Through this study, we established some objective criteria for comparison. Based on these criteria, we gave a comparative table of these different tools. We believe that research in Web mining is promising as well as challenging, and this field will help produce applications that can more effectively and efficiently utilize the Web of knowledge. We are currently working to design and implement a Web mining system based on multi-agents technology. We pretend that such system reduce the information overload and search depth. That is helpful to users using the web within a platform for ecommerce or eLearning.